\documentclass[aps,twocolumn,showpacs,preprintnumbers,amsmath,amssymb,nofootinbi
b,superscriptaddress,showkeys]{revtex4}

\usepackage{epsfig}
\usepackage{graphicx}
\begin{document}

\title{Renormalized Quarkonium}
\author{J. Segovia}\email{segonza@usal.es}
\author{D.R. Entem}\email{entem@usal.es}
\author{F. Fern\'andez}\email{fdz@usal.es}
\affiliation{Grupo de F\'{\i}sica Nuclear and IUFFyM, \\ Universidad de
Salamanca, E-37008 Salamanca, Spain}
\author{E. Ruiz Arriola}\email{earriola@ugr.es} \affiliation{Departamento de 
F\'isica At\'omica, Molecular y Nuclear, \\ Universidad de Granada, E-18071 
Granada, Spain.}
\affiliation{Instituto Carlos I de F{\'\i}sica Te\'orica y Computacional,
Universidad de Granada, E-18071 Granada, Spain}
\date{\today}

\begin{abstract} 
\rule{0ex}{3ex} We extend our previous study of the vector charmonium states
within a renormalization approach with boundary conditions to the full spectrum
of charmonium and bottomonium. On the light of the predicted spectrum we comply
to assignments suggested in the literature. A comparison with the regularized
quark model is also included.
\end{abstract}


\maketitle

%=========================================

\section{Introduction}
\label{sec:Introduction}

% QCD forces and heavy quark bound states G.S. Bali Phys. Rep. 343 1-136 (2001)

The discovery of charmonium-like states by the $B$-factories named
$XYZ$ mesons, which could not simply be described by the naive quark
model, have increased the interest for new and more accurate
charmonium potentials trying to accommodate these states.

The static charmonium potential can be parametrized phenomenologically
as a coulomb plus linear part together with spin-spin, spin-tensor and
spin-orbit terms as leading spin dependent
corrections~\cite{PhysRevD.17.3090,PhysRevD.21.203}. From a more
fundamental point of view, this potential can be derived from first
principles of QCD in the lattice with relativistic corrections
classified in powers of the inverse quark
mass~\cite{RevModPhys.77.1423}. Recently these potentials have been
improved either by new approaches in lattice
QCD~\cite{PhysRevLett.107.091601} or by matching the long range part
calculated by lattice simulations of the full QCD with results from
perturbative QCD at short range~\cite{Laschka:2012cf}. Despite
these improvements the short range part of all these potentials still
suffers from singularities which can be only handled by using {\it ad
  hoc} regulators. This triggers an unpleasant short distance
sensitivity.

In Ref.~\cite{Segovia:2011tb} we have developed an approach which
reduces the effects of the {\it ad hoc} regulators, treating exactly
the singular contributions of the potential following renormalization
ideas. Instead of fitting the regulators to reproduce the ground state
mass we use this mass as input parameter of our calculation.  The
rest of the spectrum is {\it predicted} from the orthogonality with
the ground state. The short range uncertainties are encoded in this
input and the rest of the spectrum only depends on the well
established pieces of the potential.

We have shown in Ref.~\cite{Segovia:2011tb} that this procedure
accurately reproduces the same numerical results as the standard
constituent quark model with extra regulators for the particular case
of the $J^{PC}=1^{--}$ states. In this paper we extend the calculation
to the full charmonium and bottomonium spectrum showing that for all
purposes the regulators only account for the ground states of $\bar c
c$ and $\bar b b$ systems with $J^{PC}$ quantum numbers. Therefore the
approach constitutes a useful tool to complement more fundamental
potentials. For completeness, we will also compare the results with
those of the original model (including regulators), where generally an
almost perfect agreement is found. Once these short range
uncertainties have been resolved we try to analyze which of the
currently existing $XYZ$ states could be identified as purely
quarkonium states.

The plan of the paper is as follows. Section~\ref{sec:CQM} explains briefly the
features of our constituent quark model and describes the renormalization
approach extended to all possible channels of quantum numbers.
Sections~\ref{sec:Charmonium} and~\ref{sec:Bottomonium} are devoted to comment
the spectrum of charmonium and bottomonium sectors within the renormalization
approach. Section~\ref{sec:Charmonium} includes a comparison with the original
model. We finish summarizing the work and giving some conclusions in
Section~\ref{sec:Conclusions}.

\section{Renormalized quarkonium model}
\label{sec:CQM}

The renormalization procedure we use exploits the local character of
the potentials and uses a radial regulator in terms of boundary
condition at a given cut-off radius $r_c$ which is made smaller than
any other length scale of the problem (typically $r_c \sim 0.01 {\rm
}$ is enough). Within the present context the procedure has been
explained in detail in~\cite{Segovia:2011tb} for which we refer for
further aspects and motivation. Here, we will only present its
generalization for any channel different than the $J^{PC}=1^{--}$.

The inter-quark potential we use is based on the one developed by
Vijande {\it et al.}~\cite{vijande2005constituent} which is able to
describe meson phenomenology from the light to the heavy quark
sector. It also successfully describes hadron phenomenology and
hadronic
reactions~\cite{Fernandez199235,PhysRevC.63.035207,PhysRevC.64.058201,
  PhysRevD.70.054022}. The quark model includes a screened linear
confinement potential based on unquenched lattice calculations
together with spin-dependent terms determined by perturbative
one-gluon-exchange as a Fermi-Breit type interaction. Although a
complete description of the potential and the parameter values are
given in Ref.~\cite{PhysRevD.78.114033}, it is instructive to write
down the explicit expressions of the (unregularized) One Gluon
Exchange (OGE) contribution to display its singularities explicitly.

The OGE potentials are 
\begin{eqnarray}
V_{OGE}^{C}(\vec{r}) &=&-\frac{4\alpha_{s}}{3}\frac{1}{r}, \\ 
V_{OGE}^{T}(\vec{r}) &=& \frac{1}{3}\frac{\alpha_{s}}{m^2}\frac{1}{r^3} S_{12},
\\
V_{OGE}^{SO}(\vec{r}) &=& \frac{2 \alpha_{s}}{m^2}\frac{1}{r^{3}}  \vec{L} \cdot
\vec{S} \, .  
\label{eq:OGE_unreg}
\end{eqnarray}
As we see, the tensor and spin-orbit potentials present $1/r^3$
singularities at short distances. Note that we have discarded the
Dirac delta function (which has the same dimensions as $1/r^3$) which
is traditionally included. While this may seem weird, these are
distributions around the origin which are not seen by the compact
support test functions implied by the boundary condition
regularization below the radial cut-off
radius~\cite{Segovia:2011tb}. This also applies to any derivatives of
the Dirac delta function. The result was
suggested~\cite{PavonValderrama:2005wv} and explicitly checked by
using a momentum space regularization with so-called
counterterms~\cite{Entem:2007jg}.

Once the constituent quark model we use has been presented, we will
apply the renormalization with boundary conditions to eliminate the
regulators of the model and treat exactly the singular contribution of
the potential. This scheme has been explained in detail
in~\cite{Segovia:2011tb} and we will provide here only the most
relevant additional aspects to deal with all $J^{PC}$ quarkonium
states.

In Ref.~\cite{Segovia:2011tb} it is shown that for unique and finite
normalizable solutions the number of free independent parameters of
the regularized theory depends on the behavior of the solution at the
origin and can be established by a simple analysis of the potentials.
The different cases that we can find are

\begin{itemize}
\item Uncoupled channels with a singular attractive potential. In this case, 
besides the normalization condition which eliminates one constant, one more
parameter is needed to renormalize the solution. If the potential is repulsive
at short range all the constants which determine the wave function are defined 
and the bound state is predicted as it usually happens for the standard non
singular quantum mechanics problem.

\item In the case of coupled channels the number of free parameters
  depends of the values of the potential near the origin, and more
  specifically on the corresponding eigenvalues of the coupled channel
  potential matrix. For two attractive potential eigenvalues we need
  three observables to fix the wave function. If there is one
  attractive eigenvalue only one parameter is needed. Finally, in the
  case of two repulsive eigenvalues the wave function is completely
  determined without any additional input parameter.

\end{itemize}

Focusing on quarkonium systems, we will work within the
nonrelativistic framework, so the dynamics of the system is given by
the Schr\"odinger equation. For a tensor and spin-orbit interaction we
find the following situations
\begin{widetext}
\begin{itemize}
\item Singlet channel ($s=0$, $l=J$)
\begin{equation}
-\frac{1}{2\mu}u''_{n}(r)+\left[V_{J,J}^{0J}(r)+\frac{J(J+1)}{2\mu r^{2}}\right]
u_{n}(r)=E_{n}u_{n}(r)
\label{eq:CE1}
\end{equation}
\item Triplet uncoupled channel ($s=1$, $l=J$)
\begin{equation}
-\frac{1}{2\mu}u''_{n}(r)+\left[V_{J,J}^{1J}(r)+\frac{J(J+1)}{2\mu r^{2}}\right]
u_{n}(r)= E_{n}u_{n}(r)
\label{eq:CE2}
\end{equation}
\item Triplet coupled channel ($s=1$, $l=J\pm1$)
\begin{equation}
\begin{split}
-\frac{1}{2\mu}u''_{n}(r)+\left[V_{J-1,J-1}^{1J}(r)+\frac{(J-1)J}{2\mu
r^{2}}\right]u_{n}(r)+V_{J-1,J+1}^{1J}(r)w_{n}(r) &= E_{n}u_{n}(r), \\
-\frac{1}{2\mu}w''_{n}(r)+V_{J+1,J-1}^{1J}(r)u_{n}(r)+\left[V_{J+1,J+1}^{1J}
(r)+\frac{ (J+1)(J+2)}{2\mu r^{2}}\right] w_{n}(r) &= E_{n} w_{n}(r),
\label{eq:CE3}
\end{split}
\end{equation}
\end{itemize}
\end{widetext}
where the energy is defined with respect to the $Q-\bar{Q}$ threshold,
$E_{n}=M_{n}-m_{Q}-m_{\bar Q}=M_{n}-2m$.

As mentioned, in order to determine the number of independent
constants we have to study the potential at short distances for
different channels.  In the $r \to 0$ limit the dominant contributions
are the tensor and the spin-orbit terms of the OGE potential. We need
to know their character for the different cases:
\begin{itemize}
\item Singlet channel ($s=0$, $l=J$). We have $\left\langle\right.\!\!
  S_{12} \!\!\left.\right\rangle=0$ and $\left\langle\right.\!\!
  \vec{L}\cdot\vec{S} \!\!\left.\right\rangle=0$, so the potential is
  regular and there is no observable free parameter. For the
  $^{1}S_{0}$ channel we have to take into account that in this case
  the potential has an attractive $\delta$-function and so is
  singular, therefore we have to set an observable for a regularized
  solution.
\item Triplet uncoupled channel ($s=1$, $l=J$). We have
  $\left\langle\right.\!\! S_{12} \!\!\left.\right\rangle=+2$ and
  $\left\langle\right.\!\! \vec{L}\cdot\vec{S}
  \!\!\left.\right\rangle=-1$, so the potential is singular attractive
  and an observable must be fixed for a regularized solution.
\item Triplet coupled channel ($s=1$, $l=J\pm1$). If we denote $l=J-1$ and
$l'=J+1$, we will have
\begin{itemize}
 \item $\left\langle\right.\!\! ^{3}l_{J}|S_{12}|^{3}l'_{J}
\!\!\left.\right\rangle=\frac{6\sqrt{J(J+1)}}{2J+1}$,
 \item $\left\langle\right.\!\!  ^{3}l_{J}|S_{12}|^{3}l_{J}
\!\!\left.\right\rangle=-\frac{2(J-1)}{2J+1}$,
 \item $\left\langle\right.\!\! ^{3}l'_{J}|S_{12}|^{3}l'_{J}
\!\!\left.\right\rangle=-\frac{2(J+2)}{2J+1}$,
 \item $\left\langle\right.\!\! ^{3}l_{J}|\vec{L}\cdot\vec{S}|^{3}l_{J}
\!\!\left.\right\rangle=J-1$,
 \item $\left\langle\right.\!\! ^{3}l'_{J}|\vec{L}\cdot\vec{S}|^{3}l'_{J}
\!\!\left.\right\rangle=-(J+2)$,
\end{itemize}
and diagonalizing the potential matrix the eigenvalues are
\begin{equation}
E(J)=-10\pm6\sqrt{1+J+J^{2}}
\end{equation}
with $J\geq1$, so we always have one negative eigenvalue and it requires one
parameter.
\end{itemize}

In order to describe a bound state we seek normalizable solutions
\begin{eqnarray}
\int_0^\infty \left[u(r)^2+w(r)^2\right]=1, 
\end{eqnarray}
where $w(r)=0$ for the uncoupled channels. This imposes conditions on the wave
functions both at infinity as well as at the origin.

Moreover, the set of equations~(\ref{eq:CE1}, \ref{eq:CE2}, \ref{eq:CE3}) must 
be accompanied by asymptotic conditions at infinity. Once we have discard the
irregular function at long distances, the wave functions at infinity have the
following behavior
\begin{eqnarray}
u(r) &\rightarrow& A_{J-1}\, e^{-\epsilon\,r}, \nonumber \\
w(r) &\rightarrow& A_{J+1}\, e^{-\epsilon\,r},
\label{eq:ABC}
\end{eqnarray}
where $A_{J-1}$ is the normalization factor and the asymptotic
$J+1/J-1$ ratio parameter is defined by
$\eta=A_{J+1}/A_{J-1}$. Ideally, one would integrate the Schr\"odinger
equation taking its solutions at infinity, Eq.~(\ref{eq:ABC}), which
depend on the binding energy and $\eta$. The singular structure of the
problem at short distances requires a specific analysis of the coupled
equations as it has been done extensively
elsewhere~\cite{Cordon:2009pj} and we adapt here for our particular
situation. The result amounts to integrate from infinity for the
physical value of $M_{0}$ and $\eta$ (or $M_{0}$ in the case of
singlet channels). Generally, the solutions diverge strongly at the
origin, so that the normalization of the state is precluded. However,
there is a particular value of $\eta$ which guarantees that the wave
function becomes normalizable~\footnote{Thus, if one imposes the
  regularity condition at the origin one will determine $\eta$ and
  therefore the wave function of the bound state. In practice,
  however, the converging solution is rather elusive since
  integrated-in solutions quickly run into diverging solution due to
  the round-off errors and dominate over the converging
  solution.}. The rest of the spectrum is then built by imposing
orthogonality of states in coupled channels (see~\cite{Segovia:2011tb}
for details in the $J/\psi$ case).

\section{Phenomenology of Charmonium states}
\label{sec:Charmonium}

The charmonium spectrum consists of eight narrow states below the open-charm
threshold $(3.73\,{\rm GeV})$ and several tens of states above the threshold,
some of them wide, because they decay into charmed mesons, some of them still
narrow, because their decay to open-charm is forbidden by some conservation
rule. Below the threshold all states are well established. Above threshold,
however there are new charmonium-like states that are very difficult to
accommodate theoretically.

In Tables $4$ and $9$ of Ref.~\cite{springerlink:10.1140/epjc/s10052-010-1534-9}
the updated new {\it conventional} and {\it unconventional} states in the
$c\bar{c}$, $b\bar{c}$ and $b\bar{b}$ sectors are given. If we focus on the $c\bar{c}$
region, there are three new states which have been recognized as $q\bar{q}$
pairs in the PDG~\cite{PDG2010} during the last years. They are the $h_{c}$
which is the $^{1}P_{1}$ state of charmonium, singlet partner of the long-known
$\chi_{cJ}$ triplet $^{3}P_{J}$ states, the $\eta_{c}(2S)$ which is the first
radial excitation of the pseudoscalar ground state $\eta_{c}(1S)$ and the
$Z(3930)$ whose assignment as the $2^{3}P_{2}$ state, $\chi_{c2}(2P)$, seems
widely accepted. The rest of the resonances of this sector, namely the
$X(3872)$, $X(3915)$, $Y(3940)$, $X(3940)$ still lack for a clear assignment.

The results of the study of the full charmonium spectrum up to total spin $J=2$ 
and for the first radial excitations within the renormalization with boundary 
conditions scheme are
shown in Table~\ref{tab:predmassescc}. In the following we will discuss our
predictions for the different channels except the $J^{PC}=1^{--}$ which has been
extensively discussed in Ref.~\cite{Segovia:2011tb} already. 

In columns four and five we compare the calculated masses within the
renormalization approach with the experimental data. As we see, once
the experimental value of the ground state mass is taken as input in
the calculation, the rest of the spectrum is reproduced fairly well.

In order to size the relevance of form factors in the excited spectrum
we also show in columns six and seven the comparison with the original
model including form factors~\cite{vijande2005constituent} and taking
the predicted mass as a parameter of the renormalization approach. The
observed tiny deviations are below a few MeV corresponding to a
marginal influence of the form factors on the excited states.  These
results actually point to the idea that the gluonic regulators are
fitted just to provide the ground state energies. Once we have
eliminated possibles bias due to the use of regulators, we are in a
favorable position to discuss possible $q\bar{q}$ assignments for the
old and new states.

\begin{table}[t!]
\begin{center}
\begin{tabular}{ccc|cc|cc} 
\hline
\hline
Particle & $J^{PC}$ & $nL$ &  REN1 & EXP. & REN2 & CQM \\
& & & (MeV) & (MeV) & (MeV)& (MeV) \\
\hline
$\eta_{c}$ & $0^{-+}$ & $1S$   & input & $2980.3\pm1.2$ & input& $2991$\\
& & $2S$   & $3634$ & $3637\pm4$& $3640$& $3643$ \\
& & $3S$   & $4046$ & & $4050$ & $4054$ \\[2ex]
$\chi_{c0}$ & $0^{++}$ & $1P$ & input &
$3414.75\pm0.31$ & input & $3452$   \\
& & $2P$ & $3872$ & $3915\pm3\pm2$~\cite{PhysRevLett.96.082003}& $3909$ & $3910$ 
\\
& & $3P$ & $4209$ & & $4243$ & $4242$ \\[2ex]
$h_{c}$ & $1^{+-}$ & $1P$ & $3516$ & $3525.42\pm0.29$& $3516$ & $3515$  \\
& & $2P$ & $3957$ & &  $3957$ & $3956$ \\
& & $3P$ & $4279$  &  &  $4279$ & $4278$ \\[2ex]
$\psi$ & $1^{--}$ & $1S$ & input & $3096.916\pm0.011$ & input& $3096$
 \\
& & $2S$ & $3704$ & $3686.093\pm0.034$ & $3703$ & $3703$  \\
& & $1D$ & $3796$ & $3775.2\pm1.7$ & $3796$ & $3796$   \\
& & $3S$ & $4098$ & $4039\pm1$ & $4097$ & $4097$  \\
& & $2D$ & $4152$ & $4153\pm3$ & $4153$ & $4153$  \\
& & $4S$ & $4390$ & $4361\pm9\pm9$~\cite{PhysRevLett.99.142002}& $4389$ & $4389$ 
\\
& & $3D$ & $4425$ & $4421\pm4$& $4426$ & $4426$  \\
& & $5S$ & $4615$ &
$4634^{+8+5}_{-7-8}$~\cite{PhysRevLett.101.172001}& $4614$ & $4614$  \\
& & $4D$ & $4640$ &
$4664\pm11\pm5$~\cite{PhysRevLett.99.142002}& $4641$ & $4641$  \\[2ex]
$\chi_{c1}$ & $1^{++}$ & $1P$ & input &
$3510.66\pm0.07$ & input & $3504$  \\
& & $2P$ & $3955$ & & $3947$ & $3947$  \\
& & $3P$ & $4278$ & &$4272$ & $4272$  \\[2ex]
$\eta_{c2}$ & $2^{-+}$ & $1D$& $3812$ & & $3812$ & $3812$  \\
& & $2D$ &$4166$& $4156^{+25}_{-20}\pm 15$ & $4166$ & $4166$  \\
& & $3D$ &$4437$& & $4437$ & $4437$  \\[2ex]
$\chi_{c2}$ & $2^{++}$ & $1P$ & input&
$3556.20\pm0.09$ &input & $3531$   \\
& & $2P$ & $3974$ & $3929\pm5\pm2$ & $3968$ & $3969$  \\
& & $1F$ & $4043$ & & $4043$ & $4043$  \\[2ex]
$\psi_{2}$ & $2^{--}$ & $1D$ & input & & $3810$ & $3810$  \\
& & $2D$ & $4164$ & & $4164$ & $4164$ \\
& & $3D$ & $4436$ & & $4436$ & $4436$ \\[2ex]
\hline
\hline
\end{tabular}
\caption{\label{tab:predmassescc} Masses, in MeV, of charmonium states. $nL$
labels the radial and angular momentum quantum numbers. When partial waves
are coupled they refer to the dominant component. We
compare with the well established states in Ref.~\cite{PDG2010} and assign
possible $XYZ$ mesons.}
\end{center}
\end{table}

\subsection{$\eta_{c}$ and $\eta_{c2}$ states}

An $\eta_{c}(1S)$ candidate has been observed thirty years ago by
CBAL~\cite{PhysRevLett.45.1150} and MARK II~\cite{PhysRevLett.45.1146}
Collaborations with a mass measurement very close to the updated world average
$2980.3\pm1.2$. In the renormalization scheme this state is taken as a
parameter and in essence fixes the spin-spin contact term interaction.

The search for a reproducible $\eta_{c}(2S)$ signal has a long 
history. Recently, Belle~\cite{PhysRevLett.89.102001} found a signal in
$B\rightarrow K\eta_{c}(2S)$ in the exclusive $\eta_{c}(2S)\rightarrow
K_{S}^{0}K^{-}\pi^{+}$ decay mode (a favorite all-charged final state for
$\eta_{c}(1S)$), at $3654\pm6\pm8\,{\rm MeV}$. Since then measurements of
$\eta_{c}(2S)$ in that mass region have been reported by
BaBar~\cite{PhysRevLett.92.142002}, CLEO~\cite{PhysRevLett.92.142001}, and
Belle~\cite{Nakazawa:2008zz} in $\gamma\gamma$-fusion to $K\bar{K}\pi$ final
states and by BaBar~\cite{PhysRevD.72.031101} and
Belle~\cite{PhysRevLett.98.082001} in double charmonium production.

Our predicted mass for the $\eta_{c}(2S)$ is $3634\,{\rm MeV}$ in very good
agreement with the updated world average reported in Ref.~\cite{PDG2010}.
Nothing is known of the next excitation, $\eta_{c}(3S)$. Our prediction is
around $4.05\,{\rm GeV}$.

The potential at short distances is regular for the $\eta_{c2}$ states.
Therefore the mass of the ground state for the $\eta_{c2}$ meson is not a
parameter and we predict $3812$, $4166$ and $4437\,{\rm MeV}$ for the
ground state and the first two radial excitations. 

We can clearly identify a second state $\eta_{c2}(2S)$ with mass $M=4166\,{\rm
MeV}$ and width $\Gamma=122.9\,{\rm MeV}$ with the resonance recently reported
by Belle at $M=4156^{+25}_{-20}\pm15\,{\rm MeV}$ with a width
$\Gamma=139^{+111}_{-61}\pm21\,{\rm MeV}$~\cite{PhysRevLett.100.202001} in the
$e^{+}e^{-}\rightarrow D^{\ast}{\bar{D}^\ast}J/\psi$. The decay of $\eta_{c2}$
to $D\bar{D}$ is forbidden being the $X(4160)\rightarrow D^{\ast}\bar{D}^{\ast}$
the most favored decay channel as shown by the data.

% We can clearly identify the second state with the resonance recently reported by
% Belle \cite{Bx} at $M=4156\pm 25\pm 15$ Mev with a width of $\Gamma=139\pm 61
% \pm 25$ in the reaction $e^+e^-\rightarrow J/\psi D^*\bar{D}^*$. The decay of
% $\eta_{c2}$ to $D\bar{D}$ is forbidden so this assignment
% justify the $D^*\bar{D}^*$ decay channel

\subsection{$h_{c}$ and $\chi_{cJ}$ states}

Two experiments reported the observation of the $h_{c}(1P)$ in $2005$. 
CLEO~\cite{PhysRevD.72.092004,PhysRevLett.95.102003} obtained a $6\sigma$
statistical significance in the isospin-forbidden decay chain
$e^{+}e^{-}\rightarrow\psi(2S)\rightarrow\pi^{0}h_{c}$, $h_{c}
\rightarrow\gamma\eta_{c}(1S)$. E835~\cite{PhysRevD.72.032001} found a
$3\sigma$ evidence in $p\bar{p}\rightarrow h_{c}$, 
$h_{c}\rightarrow\gamma\eta_{c}(1S)$, $\eta_{c}(1S)\rightarrow\gamma\gamma$.

The precision measurement of its mass was reported by CLEO in
$2008$~\cite{PhysRevLett.101.182003}, $3525.28\pm0.19\pm0.12)\,{\rm MeV}$. Later
BES III~\cite{PhysRevLett.104.132002} has confirmed this with a mass of
$3525.40\pm0.13\pm0.18\,{\rm MeV}$. It was important to measure the mass of this
state because the spin-averaged centroid of the triplet states
\begin{equation}
\left\langle m(1^{3}P_{J}) \right\rangle
\equiv \frac{m_{\chi_{c0}}+3m_{\chi_{c1}}+5m_{\chi_{c2}
}}{9},
\end{equation}
is expected to be near the $h_{c}(1P)$ mass. The lattice data show a vanishing
long-range component of the spin-spin potential. Thus, the potential appears to
be entirely dominated by its short-range, delta-like, part, suggesting that the
$^{1}P_{1}$ should be close to the center-of-gravity of the $^{3}P_{J}$ system.
So it makes the hyperfine mass splitting, $\Delta m_{hf}[h_{c}(1P)]=\left\langle
m(1^{3}P_{J})\right\rangle-m[h_{c}(1P)]$ an important measurement of the
spin-spin interaction.

The centroid of the $1^3P_J$ states $(\chi_{0,1,2})$ is known to
be~\cite{PDG2010} $3525.30\pm0.04\,{\rm MeV}$ and then the hyperfine splittings
are $+0.02\pm0.23\,{\rm MeV},$ from CLEO and $-0.10\pm0.22~\mathrm{MeV}$ from
BES III.

Table~\ref{tab:predmassescc} shows the masses for three radial excitations of
the singlet $^{1}P_{1}$ and the triplet $^{3}P_{J}$ mesons. In
Table~\ref{tab:centroidcc} we show the comparison between the centroid of
$^{3}P_{J}$ states and the corresponding $h_{c}$ mass for the ground state and
the first radial excitation, showing that our spin-spin interaction is
negligible for these channels and it is in perfect agreement with the lattice
expectations and the experimental measurements for the ground state.

\begin{table}[t!]
\begin{center}
\begin{tabular}{c|cc|cc|c}
\hline
\hline
& \multicolumn{2}{|c|}{CQM} & \multicolumn{2}{|c|}{REN.} & \\
\hline
nL & $M(h_{c})$ & $C_{\rm the}$ & $M(h_{c})$ & $C_{\rm the}$ &
$C_{\rm exp}$ \\
% & (MeV) & (MeV) & (MeV) & (MeV) & (MeV) \\
\hline
$1P$ & $3515$ & $3513$ & $3516$ & $3513$ & $3525.30\pm0.20$ \\
$2P$ & $3956$ & $3955$ & $3957$ & $3955$ & \\
\hline
\hline
\end{tabular}
\caption{\label{tab:centroidcc} The theoretical masses, in MeV, of the first
two radial excitations of $h_{c}$ compared with the spin-averaged centroid, in
MeV, of the triplet states. We compare with the experimental
data~\cite{PDG2010}.}
\end{center}
\end{table}

The mean $2P$ multiplet mass is predicted to be near $3.95\,{\rm GeV}$. Although
no $2P$ $c\bar{c}$ states have been clearly seen experimentally, there are a lot
of states reported from the different Collaborations which claim enhancements in
that energy region. Concerning the rest of the new states, none of them appears
in our calculation as a $c\bar c$ state except the $Z(3930)$. This fact agrees
with the results of a recent calculation which shows that these states can be
interpreted as molecular resonances~\cite{ortega2012molecular}.

The $Z(3930)$ was reported by Belle in $\gamma \gamma \rightarrow D\bar{D}$ with
a mass and width $M=3929\pm5\pm2\,{\rm MeV}$ and $\Gamma=29.9\pm10\pm2\,{\rm
MeV}$~\cite{PhysRevLett.96.082003}. The two photon width is measured to be
$\Gamma_{\gamma\gamma}B(Z(3930)\rightarrow D\bar{D})=0.18\pm0.05\pm0.03\,{\rm
keV}$. Moreover the $D\bar{D}$ angular distribution is consistent with $J=2$.
The $\chi_{c2}(2P)$ state is a good candidate for the $Z(3930)$. We obtained
a mass of $3974\,{\rm MeV}$ and the total width $\Gamma=49.1\,{\rm MeV}$ is
comparable with the experimental data. Finally the experimental two photon width
compares nicely with our result $\Gamma_{\gamma\gamma}B(Z(3930)\rightarrow
D\bar{D})=0.15\,{\rm keV}$.

\begin{table}[t!]
\begin{center}
\begin{tabular}{ccclc}
\hline
\hline
Particle & $J^{PC}$ & $nL$  & REN1 & EXP. \\
& &  & (MeV) & (MeV) \\
\hline
$\eta_{b}$ & $0^{-+}$ & $1S$  & input & $9390.9\pm2.8$ \\
& & $2S$  & $9957$ & $9999.0 \pm 3.5^{+2.8}_{-1.9}$~\cite{arXiv12056351} \\
& & $3S$  & $10306$ & \\[2ex]
$\chi_{b0}$ & $0^{++}$ & $1P$  & input &
$9859.44\pm0.42\pm0.31$ \\
& & $2P$ & $10226$ & $10232.5\pm0.4\pm0.5$ \\
& & $3P$  & $10505$ & \\[2ex]
$h_{b}$ & $1^{+-}$ & $1P$ & $9879$ & $9898.25\pm1.06^{+1.03}_{-1.07}$ \\
& & $2P$  &$10241$& $10259.76\pm0.64^{+1.43}_{-1.03}$ \\
& & $3P$  & $10516$  & \\[2ex]
$\Upsilon$ & $1^{--}$ & $1S$ & input & $9460.30\pm0.26$ \\ 
& & $2S$  & $9992$ & $10023.26\pm0.31$ \\
& & $1D$  & $10117$ & \\
& & $3S$  & $10331$ & $10355.2\pm0.5$ \\
& & $2D$  & $10414$ & \\
& & $4S$  & $10592$ & $10579.4\pm1.2$ \\
& & $3D$  & $10653$ & \\
& & $5S$  & $10805$ & $10865\pm8$\\
& & $4D$  & $10853$ & \\
& & $6S$ & $10984$ & $11019\pm8$ \\
& & $5D$  & $11023$ & \\[2ex]
$\chi_{b1}$ & $1^{++}$ & $1P$  & input & $9892.78\pm0.26\pm0.31$ \\
& & $2P$  & $10254$ & $10255.46\pm0.22\pm0.50$ \\
& & $3P$  & $10527$ & \\[2ex]
$\eta_{b2}$ & $2^{-+}$ & $1D$  & $10123$ & \\
& & $2D$  & $10419$  & \\
& & $3D$  & $10658$  & \\[2ex]
$\chi_{b2}$ & $2^{++}$ & $1P$  & input &
$9912.21\pm0.26\pm0.31$ \\
& & $2P$ & $10248$ & $10268.65\pm0.22\pm0.50$ \\
& & $1F$  & $10315$ & \\[2ex]
$\Upsilon_{2}$ & $2^{--}$ & $1D$  & input & $10163.7\pm1.4$ \\
& & $2D$  & $10418$ & \\
& & $3D$ & $10657$ & \\[2ex]
\hline
\hline
\end{tabular}
\caption{\label{tab:predmassesbb} Masses in MeV of bottomonium states. The
notation for the quantum numbers is the same as in Table~\ref{tab:centroidcc}. We
compare with the well established states in Ref.~\cite{PDG2010}.}
\end{center}
\end{table}

\section{Bottomonium}
\label{sec:Bottomonium}

In this Section we extend the renormalization with boundary conditions scheme to
the study of the bottomonium states. Although the $B$-factories are not usually
considered as ideal facilities for the study of the bottomonium spectrum since
their energy is tuned to the peak of the $\Upsilon(4S)$ resonance, which decays
in almost $100\%$ of cases to a $B\bar{B}$ pair, BaBar and Belle have reported
data samples at various energies in the bottomonium region that made possible
discoveries like the $\eta_{b}$~\cite{PhysRevLett.101.071801}, $h_{b}(1P)$ and
$h_{b}(2P)$~\cite{Adachi:2011ji}. Nevertheless the number of states is lower 
than in the charmonium spectrum.

As in the previous Section, column four of Table~\ref{tab:predmassesbb} shows
the calculated masses within the renormalization approach up to total spin $J=2$
and for the first radial excitations. As in the case of the charmonium
spectrum, when applicable, we indicate the fact that the ground state mass is
used as an input parameter.

In the $J^{PC}=1^{--}$ channel there are less data for the bottomonium
spectrum than for the charmonium.  The $S$ and $D$ wave states are
almost degenerate for the highest excited states. In the charmonium
these doublets has been measured and we can assign them to $S$ and
$D$ waves. In the bottomonium the splittings are smaller and
experimentally they have not been resolved.  We follow the assignment
of the Particle Data Group up to $4S$ states and assumed that the
others are $S$ wave. This is the reason why we assigned the
experimentally measured state at $M=11019$ MeV to our state
$6S$ and not to our $5D$.

As a general trend the experimental data are well reproduced by the
calculation and a number of excited states are given which can be
tested at LHCb and other $B$-factories. 
The $\eta_{b}(2S)$  has been very recently measured
by the Belle Collaboration~\cite{arXiv12056351} with a mass
$M=9999.0\pm 3.5^{+2.8}_{-1.9}$ which is in reasonable agreement with our 
$M=9957$ MeV result.

The hyperfine mass-splitting of singlet-triplet states, ${\Delta
m}_{hf}[\eta_{b}(1S)] = m(\Upsilon(1S))-m(\eta_b(1S))$, probes the spin-dependence
of bound-state energy levels, and, once measured, imposes constraints on
theoretical descriptions. It is given experimentally by
\begin{equation}
{\Delta m}_{hf}[\eta_{b}(1S)]=69.6\pm2.9\,{\rm MeV}
\end{equation}

In the renormalization scheme this splitting is fixed and we can only
predict the splittings for radial excitations which has not yet been 
measured.

\begin{table}[t!]
\begin{center}
\begin{tabular}{c|cc|cc|c}
\hline
\hline
& \multicolumn{2}{|c|}{CQM} & \multicolumn{2}{|c|}{REN.} & \\
\hline
nL & $M(h_{b})$ & $C_{\rm the}$ & $M(h_{b})$ & $C_{\rm the}$ & $C_{\rm exp}$ \\
% & (MeV) & (MeV) & (MeV) & (MeV) & (MeV) \\
\hline
$1P$ & $9879$ & $9879$ & $9879$ & $9879$ & $9899.87\pm0.27$ \\
$2P$ & $10240$ & $10240$ & $10241$ & $10240$ & $10260.24\pm0.36$ \\
\hline
\hline
\end{tabular}
\caption{\label{tab:centroidbb} The theoretical masses, in MeV, of the first
two radial excitations of $h_{b}$ compared with the spin-averaged centroid, in
MeV, of the triplet states. We compare with the experimental
data~\cite{PDG2010}.}
\end{center}
\end{table}

In the case of the centroid of the $\chi_{bJ}(nP)$ states with $n=1,\,2$ is known to be~\cite{PDG2010}
$9899.87\pm0.27\,{\rm MeV}$ and $10260.24\pm0.36\,{\rm MeV}$, respectively. The
hyperfine splittings measured by the Belle Collaboration~\cite{Adachi:2011ji}
are $\Delta m_{hf}[h_{b}(1P)]=+1.6\pm1.5\,{\rm MeV}$ and $\Delta
m_{hf}[h_{b}(2P)]=+0.5^{+1.6}_{-1.2}\,{\rm MeV}$ which are compatible with zero.

Table~\ref{tab:predmassesbb} shows the masses for three radial excitations of
the singlet $h_b$ and the triplet $\chi_{bJ}$ mesons. They are in
reasonable agreement with the experimental data. In Table~\ref{tab:centroidbb}
we show the comparison between the centroid of $\chi_{bJ}$ states and the
corresponding $h_{b}$ mass for the ground state and the first radial excitation,
showing that our spin-spin interaction is negligible.

$\chi_{bJ}(1P)$ and $\chi_{bJ}(2P)$ with $J=2,1,0$ were discovered earlier in
1982~\cite{PhysRevLett.49.1612,PhysRevLett.49.1616} and
1983~\cite{PhysRevLett.51.160,Pauss1983439}, respectively. Their masses have not
changed much since then and our theoretical prediction through both schemes are
very close to the experimental values.

\section{Conclusions}
\label{sec:Conclusions}

Based on an earlier work, we have reanalyzed the calculation of the charmonium
spectrum in a constituent quark model using a renormalization scheme with
boundary conditions. This approach avoids explicitly the introduction of
phenomenological form factors taking as a parameter the mass of the ground
state. Thus, the only relevant physical information on the form factors is to
tune the value of the ground state energy. Once this fact has been established
we have applied the renormalization framework to provide some basis to ``bare''
$q\bar{q}$ assignments of mesonic states.

We obtain a spectrum in reasonable agreement with the experimentally well
established data. For instance, we obtain $\Delta m_{hf}[h_{c}(nP)]$ compatible 
with zero due to the fact that the hyperfine contact term has
been included in the renormalization conditions. We also assign 
certain $XYZ$ mesons according to our model.

For the phenomenologically successful model of
Ref.~\cite{vijande2005constituent} where {\it ad hoc} form factors are
introduced as regulators, we find an almost perfect agreement with the
renormalization approach. This result provides confidence on the way the
original model took into account the unknown short distance dynamics. In
addition, we have extended this study to the bottomonium sector obtaining
similar conclusions as in the charmonium sector.

\vspace*{0.5cm}

\begin{acknowledgments}

This work has been partially funded by Ministerio de Ciencia y Tecnolog\'ia
under Contract No. FPA2010-21750-C02-02, by the European Community-Research
Infrastructure Integrating Activity 'Study of Strongly Interacting Matter'
(HadronPhysics3 Grant No. 283286), the Spanish Ingenio-Consolider 2010 Program
CPAN (CSD2007-00042), the Spanish DGI and FEDER funds with grant no.
FIS2011-24149, and by Junta de Andaluc{\'\i}a grants no. FQM225.

\end{acknowledgments}

\bibliography{renormalized_Quarkonium1}

\end{document}